\documentclass[conference,a4paper]{IEEEtran}
\IEEEoverridecommandlockouts
\usepackage{cite}
\usepackage{amsmath,amssymb,amsfonts}
\usepackage{algorithmic}
\usepackage{graphicx}
\usepackage{textcomp}
\usepackage{xcolor}
\def\BibTeX{{\rm B\kern-.05em{\sc i\kern-.025em b}\kern-.08em
    T\kern-.1667em\lower.7ex\hbox{E}\kern-.125emX}}

\usepackage[hidelinks]{hyperref}
    
\begin{document}

\title{Quantum Machine Learning for Climate Modelling
\thanks{This project was made possible by the DLR Quantum Computing Initiative and the Federal Ministry for Economic Affairs and Climate Action; \url{qci.dlr.de/projects/klim-qml}. V.E. was additionally supported by the Deutsche Forschungsgemeinschaft (DFG, German Research Foundation)
through the Gottfried Wilhelm Leibniz Prize awarded to Veronika Eyring (Reference No. EY 22/2-1).}
}




\author{\IEEEauthorblockN{Mierk Schwabe\IEEEauthorrefmark{1}, Lorenzo Pastori\IEEEauthorrefmark{1}, Valentina Sarandrea\IEEEauthorrefmark{1}, and Veronika Eyring\IEEEauthorrefmark{1}\IEEEauthorrefmark{2}}
\IEEEauthorblockA{\IEEEauthorrefmark{1}Deutsches Zentrum f\"ur Luft- und Raumfahrt e.V.\\
Institut f\"ur Physik der Atmosph\"are, Oberpfaffenhofen, Germany
}
\IEEEauthorblockA{\IEEEauthorrefmark{2}University of Bremen, Institute of Environmental Physics (IUP), Bremen, Germany}\\
Email: mierk.schwabe@dlr.de
}

\maketitle

\begin{abstract}
Quantum machine learning (QML) is making rapid progress, and QML-based models hold the promise of quantum advantages such as potentially higher expressivity and generalizability than their classical counterparts. Here, we present work on using a quantum neural net (QNN) to develop a parameterization of cloud cover for an 
Earth system model (ESM). ESMs are needed for predicting and projecting climate change, and can be improved in  hybrid models incorporating both traditional physics-based components as well as machine learning (ML) models. We show that a QNN can predict cloud cover with a performance similar to a classical NN with the same number of free parameters and significantly better than the traditional scheme. We also analyse the learning capability of the QNN in comparison to the classical NN and show that, at least for our example, QNNs learn more consistent relationships than classical NNs.
\end{abstract}


\begin{figure}
    \includegraphics[width=\linewidth]{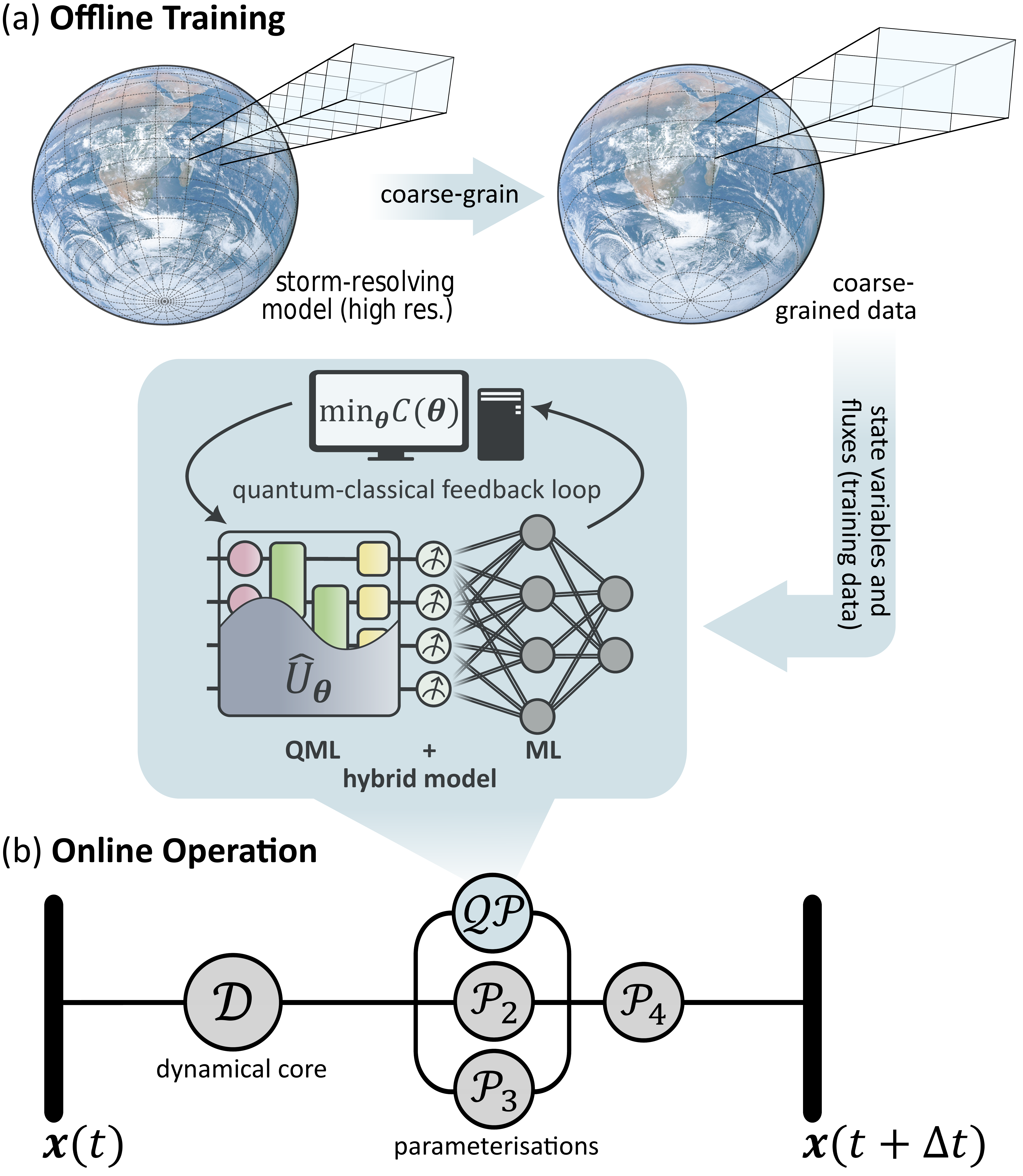}
    \caption{\label{fig:param}Hybrid quantum-classical approach for QML-based parameterisations, analogous to the approach typically taken for developing classical ML-based parameterizations. (a) Training data are generated with short high-resolution (storm-resolving) climate model runs, or with Earth observations. Next, these data are coarse-grained to the resolution of the target climate model, and the target output variables are calculated, for instance the flux due to convection. The data are then used for offline training of the QML model, typically in a hybrid fashion together with a classical computer. (b) The resulting QML model of a given process replaces a conventional parameterization and is run at every time step in the coarse climate model after the dynamical core that solves equations describing the resolved dynamics. From \cite{Schwabe2025}, CC BY license.}
\end{figure}

\section{Introduction}

\begin{figure*}[bth!]
    \includegraphics[width=\linewidth]{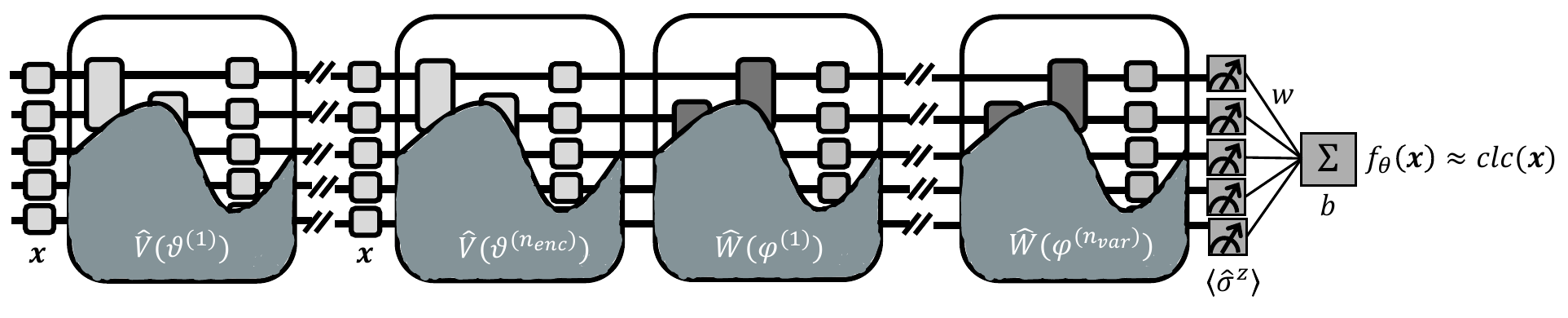}
    \caption{\label{fig:arch}Sketch of the QNN architecture. The number of used qubits (here shown schematically as 5) is equal to the number of input variables used, either 8 or 6. The data $\bf x$ are uploaded $n_\mathrm{enc}$ times as angles of single-qubit rotations, each variable to a specific qubit. These re-uploading gates are interleaved with variational blocks $\hat{V}(\boldsymbol{\vartheta}^{(k)})$ containing entangling gates and trainable parameters $\boldsymbol{\vartheta}^{(k)}$, $(k = 1, \dots , n_\mathrm{enc})$. then, another sequence of $n_\mathrm{var}$ variational blocks $\hat{W} (\boldsymbol{\varphi}^{(\ell)})$, $(\ell = 1, \dots , n_\mathrm{var})$ are applied. Finally, the expectation values of $\hat{\sigma}^z$  on all qubits are measured, and a weighted average with trainable weights $w$ and a bias term $b$ is calculated. During training, the parameters $\boldsymbol{\theta} = \{ \{ \boldsymbol{\vartheta}^{(k)} \}_k, \{ \boldsymbol{\varphi}^{(\ell)}\}_\ell,\mathbf{w},b \}$ are adjusted such that the result $f_\theta(\bf x)$ approximates the cloud cover clc($\mathbf x$). Following \cite{Pastori2025}.}
\end{figure*}

Earth system models (ESMs) are global coupled models of the different components of the Earth system and their interactions, including the atmosphere, ocean, land, and cryosphere. They are used to project future climate change for better climate change mitigation and adaptation strategies. ESMs used for projections of the future climate typically have resolutions of the order of 10s of km. Higher resolutions are possible, but are extremely computationally demanding and thus cannot be used for ensemble generation, which is needed for uncertainty quantification. Even high resolution models cannot resolve all processes, such as cloud microphysics and turbulence, and some long-standing biases remain \cite{Stevens2019b,Kodama2021}. At the typical coarse resolutions, also various further dynamical effects are not resolved, e.g., gravity waves and convection. The influences of these unresolved processes on the climate are taken into account with empirical {\itshape parameterizations}. The use of these parameterizations leads to uncertainties and biases in climate models. Machine learning (ML) has already demonstrated the capability to enhance physics-based ESMs \cite{Eyring2023}. In hybrid (ML \& physics) models individual parameterizations or combinations of various parameterizations are replaced with ML-based models \cite{Bracco2024}. The ML models can be trained on short high-resolution data or Earth observations, or combinations thereof (e.g., first training on global high-resolution simulations and then refining the model with observations (e.g., \cite{Grundner2024})). Another possibility is to replace the entire model with an ML-based emulator (e.g., \cite{Bodnar2024}). Both approaches, however, have difficulties when generalizing beyond the training regimes, for instance, in a changing climate where the mean and extremes of climate variable distributions
are shifting \cite{Gentine2021}. 

A hierarchy of climate models will continue to be required, ideally comprising hybrid models \cite{Eyring2023}. Quantum computing offers additional potential to improve and accelerate climate models. Potential application areas range from accelerating the resolution of the underlying fundamental equations to tuning the climate models (i.e., finding the best set of free parameters to match observations), to using QML approaches to replace traditional parameterizations, analogous to classical ML approaches. Several works have demonstrated good generalization capabilities of quantum neural networks \cite{Banchi2021,Caro2021,Caro2022a,Haug2024}, higher expressivity \cite{Du2020,Yu2023}, and potential good trainability \cite{Abbas2021}. Here, we explore using QML for climate modelling, in particular for a cloud cover parameterization for a climate model. 

In the typical training process for a data-driven parameterization based on high-resolution data, the training data are coarse-grained to the resolution of the target climate model (Fig.~\ref{fig:param}). This makes it possible to use the coarse-grained state variables as input to the QML model during training, serving as a proxy for the state variables in the coarse climate model. The output during the training represents the subgrid-scale process that the target parameterization models, e.g., fluxes due to unresolved convection, or the cloud cover in a given climate model cell. The data-driven parameterization thus learns these unresolved processes as a function of the coarse state of the Earth system in the given cell. After training, the QML model is then coupled to the climate model and run at every time step together with the other parameterizations after the dynamical core (representing the resolved processes). 

Bazgir and Zhang \cite{Bazgir2024} recently developed QML models to represent target values such as snow and rain rates, based on the ClimSim dataset \cite{Yu2023a}. This dataset is produced with a multiscale modeling framework (MMF) that embeds a kilometer-resolution cloud-resolving model within each atmospheric column of a coarser host climate model to replace traditional convection and cloud parameterizations \cite{Hu2024}. Bazgir and Zhang \cite{Bazgir2024} trained various QML models to represent the high-resolution embedded output as a function of the coarse state. They tested a quantum convolutional neural network, a quantum multilayer perceptron, and a quantum encoder-decoder, and compared them to their classical counterparts. They found that the quantum models typically outperformed the classical ones. 

In our work, we investigate a quantum neural network for the representation of cloud cover in a climate model.  

\section{Methods}

\subsection{QML and ML models of cloud cover}

We used high-resolution climate model data as training data \cite{Pastori2025}, in contrast to training on the MMF as above. This has the advantage that we were able to use the same climate model to generate the training data as the target model (ICON \cite{Mueller2025}), and the data are more realistic as no scale-separation is assumed a priori \cite{Beucler2024}, but the training data are less obvious to prepare. In this work, global high-resolution (2.5\,km in the horizontal) simulations with the ICON climate model from the DYAMOND project \cite{Stevens2019b} are used. As in \cite{Giorgetta2018}, in the high-resolution data, a grid cell is considered to be fully {\textit cloudy} when specific cloud condensate (water or ice) content exceeds a threshold, and cloud-free otherwise. The data are then coarse-grained to a horizontal resolution of 80\,km \cite{Grundner2022}. The cloud cover after coarse-graining is between 0 and 1 and represents the cloudy fraction of the cell. The QML models are trained to learn the cloud fraction as a function of the coarse state variables (input features for the QML model) specific humidity $q_v$, specific cloud water content $q_c$, specific cloud ice content $q_i$, air temperature $T$, pressure $p$, and magnitude of horizontal wind $h_w$, as well as geometric height $z_g$ and latitude $\phi$, or a subsection thereof (excluding height and latitude). 
A parameterized quantum circuit is used as QML model, with the number of qubits equal to the number of input features (Fig.~\ref{fig:arch}). The qubit register is initialized in the $|0\rangle$ state. After suitable rescaling, the input features $\bf x$ are encoded as angles of single-qubit rotations in multiple encoding layers $\hat{S}({\bf x})$
\begin{equation}
    \hat{S}(\mathbf{x}) = \prod_{n=1}^N \exp{-i \frac{x_n}{2} \hat{\sigma}_n^x} \equiv \hat{R}_x(\mathbf{x})
\end{equation}
with $x_n$ denoting the $n$th component of the input $\mathbf{x}$ with $N$ features, and $x$ the rotation axis. 

\begin{figure}
\centering
    \includegraphics[width=0.9\linewidth]{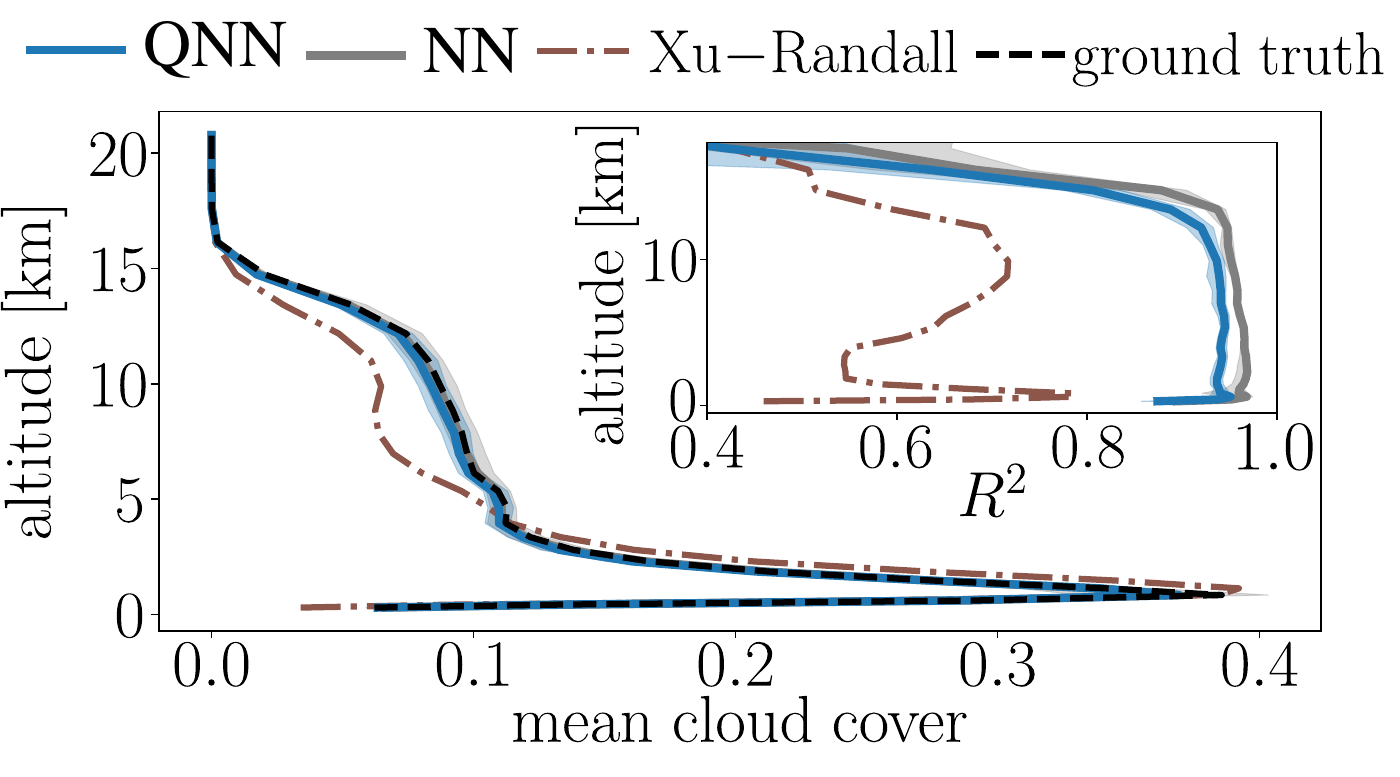}
    \caption{\label{fig:qmlcc}Mean cloud cover as a function of altitude learned by a QNN model (blue solid line) and a classical NN with a comparable number of free parameters (grey solid line), compared with the ground truth (black dashed line) and the Xu-Randall scheme typically used in climate models (brown dash-dotted line). The shaded areas correspond to the spread over an ensemble of 20 training instances. Inset: Coefficients of determination $R^2$ for the various models as a function of altitude. Adapted from \cite{Pastori2025}.} 
\end{figure}

Data reuploading \cite{PerezSalinas2020,Schuld2021} is used to increase the number of Fourier frequencies that the model is able to capture \cite{Schuld2021,Casas2023}. These encoding layers are interleaved with variational blocks $\hat{V}({\boldsymbol{\vartheta}}^{(k)})$ with the index $k$ and trainable parameters ${\boldsymbol{\vartheta}}^{(k)}$. The $\hat{V}({\boldsymbol{\vartheta}}^{(k)})$ also contain entangling operations. We tested various circuit architectures \cite{Pastori2025}. For the results presented in this paper (in blue in Figs.~\ref{fig:qmlcc}, \ref{fig:shotnoise}, \ref{fig:XAI}) 
\begin{align}
      \hat{V}(\boldsymbol{\vartheta}) &= \hat{R}_{yy}(\boldsymbol{\vartheta}_{(2N-1)\rightarrow (3N-3)}) \times \\ 
&\phantom{=.} \hat{R}_{xx}(\boldsymbol{\vartheta}_{N\rightarrow (2N-2)}) \times  \nonumber \\
&\phantom{=.} \hat{R}_{zz}(\boldsymbol{\vartheta}_{1\rightarrow(N-1)}) \nonumber 
\end{align}
with $\hat{R}_{\alpha \alpha}(\boldsymbol{\vartheta}) = \prod_{n=1}^{N-1} \exp{-i \frac{{\vartheta}_n}{2} \hat{\sigma}_n^\alpha} \hat{\sigma}_{n+1}^\alpha$, $\alpha = x, y, z$, and $\boldsymbol{\vartheta}_{i \rightarrow j}$ the slice of $\boldsymbol{\vartheta}$ from the $i$th to the $j$th component.

After $n_\text{enc}$ applications of the blocks $\hat{S}({\bf x})$ and $\hat{V}({\boldsymbol{\vartheta}}^{(k)})$, another $n_\text{var}$ variational blocks $\hat{W}({\boldsymbol{\varphi}^{(\ell)}})$ with index $\ell$ are applied to introduce additional trainable parameters ${\boldsymbol{\varphi}}^{(\ell)}$ and entangling operations. For the results in blue in Figs.~\ref{fig:qmlcc}, \ref{fig:shotnoise}, \ref{fig:XAI}
\begin{align}
\hat{W}(\boldsymbol{\varphi}) &= \hat{R}_x(\boldsymbol{\varphi}_{(3N-2)\rightarrow (4N-3)}) \times \\
&\phantom{=.} \hat{R}_{yy}(\boldsymbol{\varphi}_{(2N-1) \rightarrow (3N-3)}) \times \nonumber \\
&\phantom{=.} \hat{R}_{xx}(\boldsymbol{\varphi}_{N \rightarrow (2N-2)}) \hat{R}_{zz}(\boldsymbol{\varphi}_{1 \rightarrow (N-1)}).  \nonumber
\end{align}

After the computation, the expectation values of all Pauli $Z$ operators are measured on the output state, and their weighted average with trainable weights $\bf w$ and bias $b$ is computed and interpreted as the fractional cloud cover in the cell. The QNNs are numerically simulated in Python using the Pennylane library \cite{Bergholm2018} and optimized with JAX \cite{jax2018github}. For QNNs implemented on a physical quantum device, the parameters $\boldsymbol{\theta}$, summarizing the trainable parameters $\boldsymbol{\vartheta}, \boldsymbol{\varphi}, \mathbf w$ and $b$, are trained in a quantum-classical feedback loop \cite{McClean2016,Cerezo2021}: The QNN is run on the quantum computer for a given set of $\boldsymbol{\theta}$, and a cost function is calculated. Depending on this cost function, a new set of $\boldsymbol{\theta}$ is proposed, and the QNN is run in the next iteration with the new parameters. In our simulated training, we use the mean squared error (MSE) over the training dataset as cost function and update the parameters $\boldsymbol{\theta}$ using gradient descent. For QNNs the parameter-shift rule \cite{Mitarai2018,Schuld2019} can be used to calculate the gradients of the cost function with respect to the parameters. 

\begin{figure}
    \centering
    \includegraphics[width=0.85\linewidth]{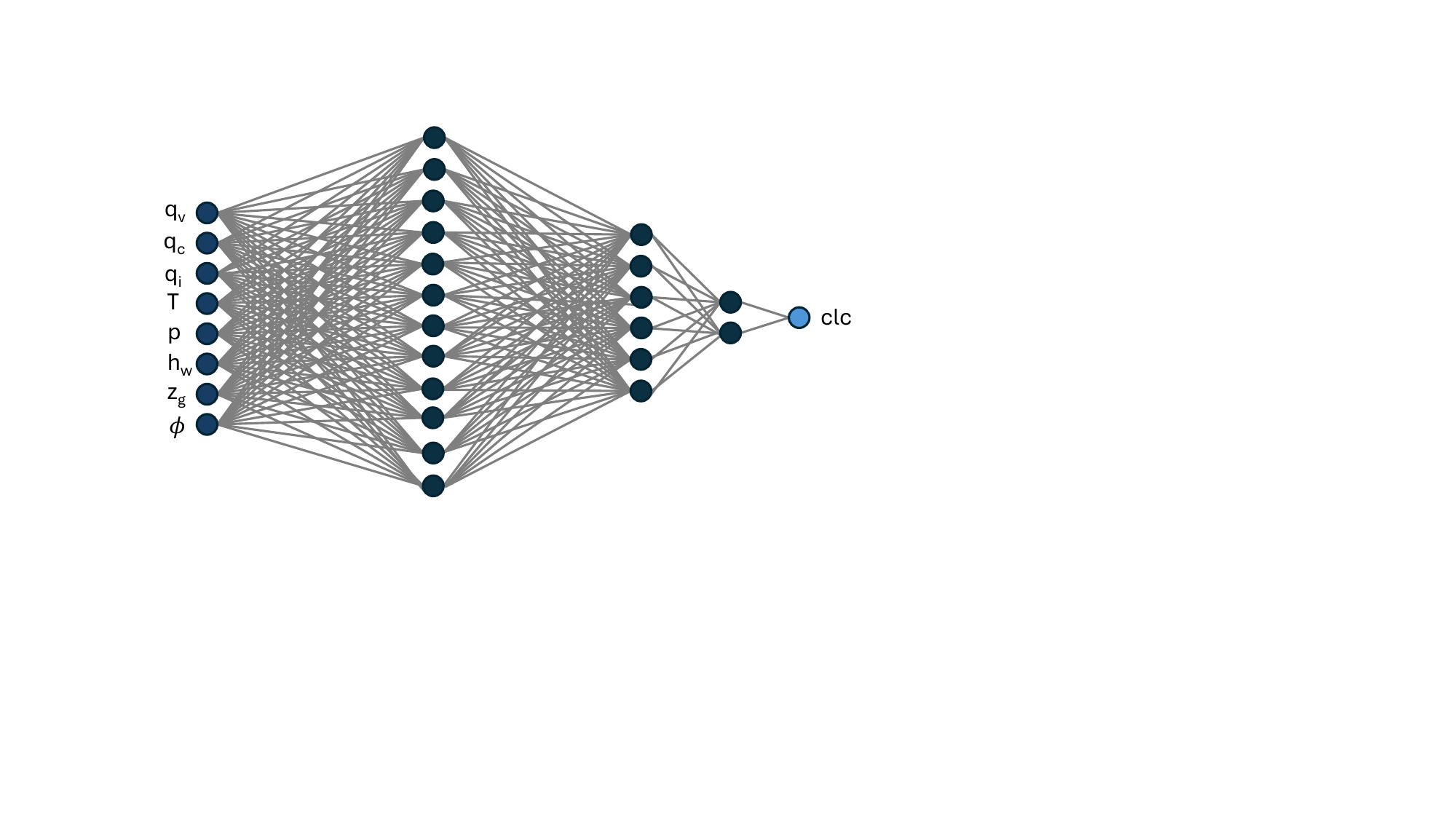}
    \caption{\label{fig:classNN}Architecture of the classical neural net with eight input features used for comparison. After the input layer there are three hidden layers with 12, 6, 2 nodes, followed by a single node for output of the predicted cloud cover.}
\end{figure}

The networks are trained for 200 epochs with $N_\mathrm{train} = 2 \times 10^5$ training data. Each epoch consists of 1000 batch updates with a batch size of 100. The QML models have 200 resp. 201 trainable parameters when using the full set of eight input features, depending on the exact architecture choices. They are compared with a classical NN (Fig.~\ref{fig:classNN}) with 203 free parameters (using eight input features). As is visible in the different architecture layouts, the classical part of the QNN is minimal: the classical operations performed at the end of the QNN correspond to a simple weighted average of the QNN measurement outputs (with trainable weights plus a bias), i.e., to a one-node linear layer in the NN terminology. The fully-classical NN used in our work instead contains multiple layers consisting of several nodes, each including a non-linear activation function (either leaky-ReLU or tanh), which is indispensable to the operation of the NN. Therefore, we ensure that the quantum part of the QNN is essential in performing the calculations.

\subsection{Shapley values for explainability}

Explainability methods for classical machine learning are already well established, even though advances in deep learning lead to a performance-explainability trade-off \cite{BarredoArrieta2020}. The application of explainability techniques to quantum machine learning is an active field of research \cite{Steinmueller2022,Power2024,GilFuster2024,Heese2025}.

In this paper, we apply a model-agnostic explainability method, namely Shapley values \cite{Shapley1953}. This method scores input features according to the
effect of adding or removing them in the output of the model \cite{GilFuster2024}. They are calculated for quantum and classical machine learning models using the SHapley Additive exPlanations (SHAP) library \cite{Lundberg2017}. 

A SHAP value $\text{shap}(x = x_0, y)$ gives the deviation in the output variable $y$ for a specific value $x_0$ of the variable $x$ from the average prediction of $y$ over a given dataset. It allows to determine the importance of the various input features for a specific output value and to interpret the relationships discovered by the data-driven model \cite{Heuer2024}. SHAP values offer both global and individual explanations of feature importance \cite{Power2024}. Specifically, in our work, we used KernelSHAP, a model-agnostic method that approximates Shapley values by treating the model as a black box and estimating contributions through perturbations of the input features \cite{Sarandrea2025}. We select 100 representative samples from the training dataset as background dataset to approximate feature attributions. Once the explainer is initialized, Shapley values are computed for the test dataset of 100,000 samples.

\section{Results}

The resulting QML model performs similarly well as a classical neural net (NN) with a similar number of free parameters, with the classical NN slightly outperforming the QNN. Both classical and quantum NN perform significantly better than the traditionally used Xu-Randall parameterization scheme \cite{Xu1996} (Fig.~\ref{fig:qmlcc}). 

The results presented so far were simulated without any noise. We also investigated the dependence of the results on shot noise. Training QNNs from the same initial parameter set, but with different shot noise realizations shows that training is successful and closely follows the noiseless training curve for a sufficiently large number of shots ($n_\text{shots} > 10^4$) \cite{Pastori2025}. When a smaller number of $n_\text{shots}$ is used, initially the mean squared error is minimized, but then the training becomes unstable. This happens when the magnitude of the gradients becomes comparable to the noise in the gradients’ evaluation. 

\begin{figure}
    \centering
    \includegraphics[width=0.9\linewidth]{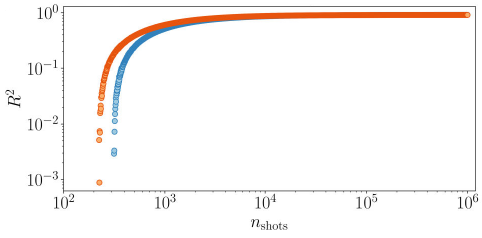}
    \caption{\label{fig:shotnoise}Test set performance of two different architectures of QNNs for cloud cover %
    using six input features 
    (blue and orange, for details see \cite{Pastori2025}) as function of the number of shots $n_\text{shots}$ used at inference. The QNNs were trained in the noiseless regime. The test set had a size of $2 \times 10^5$. Adapted from \cite{Pastori2025}.}
\end{figure}

We also analysed the performance on a test dataset with a varying number of shots $n_\text{shot}$ (Fig.~\ref{fig:shotnoise}). The performance, as measured by the coefficient of determination $R^2$, rapidly degrades for smaller $n_\text{shot}$, but is stable for $n_\text{shot} > 10^4$. Techniques such as variance regularization \cite{Kreplin2024} can be used to minimize the number of shots needed, both during and after training \cite{Pastori2025}. These results demonstrate the usefulness of quantum computing for developing parameterizations in climate models. 

Next, we investigate the learning capabilities of the QML model in comparison with the classical NN with SHAP values \cite{Sarandrea2025}. In the framework of explainability of ML models one of the first questions one may ask concerns the importance ranking of the input features in the model predictions, to see whether this ranking aligns with our physical expectations. In our case, in order to make a comparison between the quantum and classical models we use SHAP values since they are model agnostic, and their definition and computation does apply to both quantum and classical models. 

In general, the analysis reveals that both classical and quantum NN place the highest importance on temperature and specific humidity, in accordance with our physical understanding. For a more thorough discussion of the specific rankings in relation with our physical expectations see \cite{Sarandrea2025}. Here, we explore whether the feature ranking is stable across different training experiments, which can be considered a proxy for the stability of the functional input-output relationships learnt by the model. 
We train two sets of ten networks starting from randomly initialized parameters. The resulting networks have very similar performance ($R^2$ between 0.92 and 0.94 for the classical NNs and between 0.89 and 0.91 for the quantum NNs). 

\begin{figure}[t]
    \centering
    \includegraphics[width=0.8\linewidth]{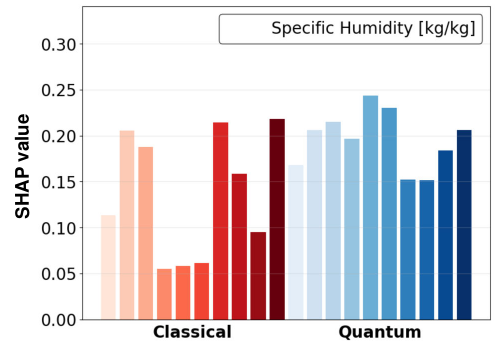}
    \caption{\label{fig:XAI}Feature importances revealed in SHAP values of specific humidity for ten instances of classical (red, left) and quantum (blue, right) neural networks. Adapted from \cite{Sarandrea2025}.}
\end{figure}

Despite the comparable performances, a SHAP analysis reveals that feature importances for the classical neural networks differ vastly among each other, while the feature importances for the quantum NNs are more stable (see Fig.~\ref{fig:XAI} for the feature importances found for specific humidity as an example, 
similar differences in variance of the feature importances are observed for other variables, especially temperature and pressure)%
. 

This means that, in this case, at least some well-performing classical neural networks learn potentially unphysical relationships -- the feature importance of specific humidity should be consistently high, since it is the most important feature determining cloud cover. The training of the quantum neural networks in the analysed example is more robust and consistent. 

\section{Conclusion and outlook}

In conclusion, we developed QML models to represent subgrid-scale cloud cover in a climate model. We showed that the QML models perform almost as well as classical neural nets with a similar number of trainable parameters and significantly better than the conventional parameterization used in climate models. We showed that, even taking into account shot noise, the QML models arrive at good predictions with a reasonable number of shots. 
We also investigated the robustness of training both the classical and the quantum models and found in limited tests that the quantum models arrive at more stable feature importances than classical models. Whether these results hold for further examples, and what the influence of the different architectures and of training procedures is, is subject of future work. 

As next step, the methods presented here could be extended to larger problem sizes, especially moving away from the cell-based approach used here by considering also neighboring cells as in \cite{Grundner2022}. Extending the work to further architectures could also prove promising, for example to quantum convolutional neural networks \cite{Cong2019} which do not suffer from problems with barren plateaus \cite{Pesah2021}. QML approaches can also be applied for more complex climate model parameterizations, such as turbulence in the boundary layer. Another interesting application would be to learn the underlying statistics in the parameterization, for instance as introduced with coarse-graining. For this, the obvious choice would be quantum generative models \cite{Amin2018,Dallaire-Demers2018,Coyle2020,Gao2022}.

These models can be developed already on current noisy intermediate-scale quantum (NISQ) devices. However, running the QML-based parameterizations coupled to the climate model requires considerable
quantum and classical runtime with overheads due to the quantum-classical coupling \cite{Schwabe2025}. While work on this coupling is ongoing, surrogates or shadows of QML models \cite{Landman2022,Schreiber2023,Jerbi2023,Sweke2025} could be used instead of running the QML models on real quantum hardware -- classical models emulating the outputs of previously trained QML models, retaining some of the potential benefits while requiring a quantum device only during the training stage. 

All these research efforts still face numerous challenges, but they also offer exciting potential to significantly boost the development of climate model parameterizations, in addition to current efforts on hybrid ML-enhanced ESMs \cite{Eyring2023} and high resolution modelling \cite{Stevens2024}. 


\bibliographystyle{IEEEtran}
\bibliography{IEEEabrv,QC4Climate,additional_refs}{}

\end{document}